\documentclass[preprint]{revtex4-1}
\usepackage{graphicx}

\newcommand{\tda}{\tilde{\delta}_a}
\newcommand{\tdo}{\tilde{\delta}_0}
\newcommand{\tq}{p}
\newcommand{\tz}{\xi}
\newcommand{\bee}{\begin{equation}}
\newcommand{\ee}{\end{equation}}
\newcommand{\beqa}{\begin{eqnarray}}
\newcommand{\eeqa}{\end{eqnarray}}
\begin{document}
\title{Repulsive Casimir Force: Sufficient Conditions}
\author{Luigi Rosa$^{1,2}$, and Astrid Lambrecht$^1$}
\address
{$^1$ Laboratoire Kastler Brossel, CNRS, ENS, UPMC-Campus Jussieu case 74,75252 Paris, France\\
$^2$ Dipartimento di Scienze Fisiche Universit\`a di Napoli Federico II , \\
and INFN, Sezione di Napoli, Monte S. Angelo, Via Cintia 80126 Napoli, Italy}
\email{ rosa@na.infn.it}
\email{astrid.lambrecht@lkb.ens.fr  }
\vskip1.5truecm
\begin{abstract}
In this paper the Casimir energy of two parallel plates made by materials of different penetration depth and no medium in between is derived. We study the Casimir force density and derive analytical constraints on the two penetration depths which are {\em sufficient conditions} to ensure repulsion. Compared to other methods our approach needs no specific model for dielectric or magnetic material properties and constitutes a complementary analysis.
\end{abstract}
\maketitle
\section{Introduction}

One of the striking features of the Casimir effect
\cite{cas48,bomu01} is the dependence of the sign of the energy, which may be
positive or negative, and of the corresponding repulsive or attractive force, on the geometry of the device and on the materials
used. The possibility of obtaining a repulsive force would open vast possibilities in the design of
MEMS and NEMS \cite{kenn02,guss06}. For example one
of the principal causes of malfunctioning in MEMS is stiction: the
collapse of nearby surfaces, resulting in their permanent adhesion.
The possibility of having a repulsive Casimir force is an
interesting way to avoid such a collapse of the structure but up to
now there are only few experimental evidences for a repulsive force
\cite{millmu,leesi,capa09}

In this paper we study the Casimir energy of two parallel
plates made by materials having different penetration depth.
A study of the Casimir force and the role of surface plasmons between dissimilar mirrors
has been carried out in the past for specific models of the dielectric and magnetic properties of
the materials \cite{astir08}. Here we propose an alternative method which is model independent and gives
thus complementary information on the possibility of repulsive Casimir forces.
We find that, depending on the relation between the penetration
depths and the distance of the plates the force can be both
attractive and repulsive. The penetration depth of materials can be
taken into account by means of its connection to the surface impedance
\cite{strat41,landau8,moste84}. The surface impedance $Z$ of any
planar surface may be defined as the ratio of the complex electric
and magnetic tangential field components at the surface
\cite{strat41}:
\bee
{\bf{E}}_t(z_0)=Z({\bf{H}}_t(z_0)\times\hat{n}) \label{eq:SI}
\ee
where $\hat{n}$ is a normal vector pointing inside the surface and
$z_0$ is the position of the surface. The main advantage of this
formula is that it relates  the tangential fields {\em outside the
material}, thus it is not necessary to consider the internal degrees
of freedom of the material which are taken into account through the
values of $Z$ \cite{strat41}. Equation (\ref{eq:SI}) can be seen as
an exact functional definition of the surface impedance so that it
can be applied to arbitrary materials \cite{esqui03} and it still
holds when a description in terms of dielectric permittivity cannot
be given \cite{geklim03}. Indeed a complete correspondence with
reflection coefficient and surface impedance  exists
\cite{strat41,esqui03}. Moreover \cite{kenn02} ``{\em for large
permeability and permittivity, the transition from attractive to
repulsive behavior depends only on the impedance $Z$}''

The paper is organized as follows: in section II the Casimir force
in the general configuration is evaluated and some limiting results
are recovered. In section III the conditions for having repulsions
are derived. Finally section IV contains remarks and conclusions.

\section{The Casimir Force}

In the following we will consider two parallel plates lying in the
$(x,y)$ plane, located at $z=0,z=a$ and characterized by different
surface impedances $Z_{(0,a)}(\omega)$ respectively. Given the
functions
$\delta_{(0,a)}(\omega)=i\frac{Z_{(0,a)}(\omega)}{\omega}$,
$Re[\delta(\omega)]$ is interpreted as the penetration depth of the
material at the frequency $\omega$ \cite{landau8,moste84} see also
\cite{geklim03}. Because of translational invariance in the $(x,y)$
plane the electric and magnetic fields can be written as (in the
following we will use natural units: $\hbar=c=1$):

\bee
{\bf{E}}({\bf{x}},t)={\bf f}(z)e^{i k_\perp\cdot x_\perp-\omega t},~~~~
{\bf B}({\bf{x}},t)={\bf g}(z)e^{i k_\perp\cdot x_\perp-\omega t}
\ee
with $k_\perp\equiv(k_x,k_y) $ and $x_\perp\equiv (x,y)$.
The Maxwell equations imply:
\bee
\frac{d^2 {\bf f}}{d z^2}+\lambda^2 {\bf f}=0;
\frac{d^2 {\bf g}}{d z^2}+\lambda^2 {\bf g}=0;
\ee
with $\lambda^2=\omega^2-k_x^2-k_y^2$. Imposing relation (\ref{eq:SI})
we obtain, for the $(x,y)$ components of ${\bf E}$ and ${\bf B}$ the following boundary
conditions \cite{moste84,geklim03}.
\bee
\left\{\begin{array}{c}
                      f_x(0)=-\delta_0(i k_x f_z(0)-f_x'(0))\\
                      f_x(a)=\delta_a(i k_x f_z(a)-f_x'(a))
                    \end{array}\right.~~~~
\left\{\begin{array}{c}
                      f_y(0)=-\delta_0(i k_y f_z(0)-f_y'(0))\\
                      f_y(a)=\delta_a(i k_y f_z(a)-f_y'(a))
                    \end{array}\right.
\ee
moreover everywhere $\nabla\cdot\bf{E}=0$ must be satisfied.

In this way, with a suitable choice of the reference frame, we find
the following dispersion equation:
\bee
\Delta_{TM}\equiv[\delta_0\delta_a( \lambda^4+k_\perp^4)+
\lambda^2(2\delta_0\delta_a k_\perp^2-1)]\sin{(a \lambda)}
-(\delta_0+\delta_a)\lambda(\lambda^2+k_\perp^2)\cos{(a \lambda)}=0
\ee
for the TM modes and
\bee
\Delta_{TE}\equiv\delta_0\delta_a (\lambda^2-1)\sin{(a \lambda)}-(\delta_0+\delta_a)
\lambda\cos{(a \lambda)}=0
\ee
for the TE ones. We use the argument theorem to obtain the Casimir
energy \cite{vanka68,bomu01} so that, after $\omega$-rotation to the
imaginary axis: $\omega\rightarrow i\zeta$, we have (for the properties of $\delta$ (or $Z$) along the imaginary axis see \cite{landau8,moste84,bara7584})
\bee
E=\frac{1}{2(2 \pi)^3}\int{d\zeta dk_x dk_y
\ln{[\Delta_{TM}(\lambda,i \zeta)\Delta_{TE}(\lambda,i \zeta)]}}
\ee
This integral diverges and, as usual, to regularize it we must
subtract the energy corresponding to the configuration with the two
plates infinitely far away ($a\rightarrow \infty$):
\beqa
\Delta_{TE}^\infty &=& -i e^{a q} \left (\frac {1} {2} {\delta_0} {\delta_a} q^2 -
\frac {{\delta_ 0} q} {2} - \frac {{\delta_a} q} {2} + \frac {1} {2} \right) \\
\Delta_{TM}^\infty &=& -\frac {i e^{a q} \left ({\delta_ 0} q^2 - q - {\delta_ 0}
          {k_x}^2 \right) \left ({\delta_a} q^2 -
     q - {\delta_a} {k_x}^2 \right)} {2 q^2}
\eeqa
with $q=\sqrt{\zeta^2+k_\perp^2}$.
Thus the renormalized Casimir energy will be given by:
\beqa
E_R &=&\frac{1}{2(2 \pi)^3}\int_{-\infty}^{\infty} d\zeta dk_x dk_y
\ln{[\left(1 - e^{-2 a q}\frac {\left (1- q\delta_a (i\zeta)\right)}
{\left (1+ q\delta_a (i\zeta)\right)}\frac{\left (1 - q\delta_0 (i\zeta) \right)}{\left (1 + q\delta_0 (i\zeta)\right)}\right)]} + \nonumber \\
 & &\ln{[\left(1 - e^{-2 a q}\frac {\left (q - \delta_a (i\zeta)\zeta^2\right)}
{\left (q + \delta_a (i\zeta)\zeta^2\right)}\frac{\left (q - \delta_0 (i\zeta)\zeta^2 \right)}{\left (q + \delta_0 (i\zeta)\zeta^2 \right)}\right)]}
\eeqa
or, in dimensionless variables
\beqa
E_R &=& \frac{1}{4 \pi^2 a^3}\int_0^\infty d\xi\int_0^\infty k_\perp d\tilde{k}_\perp
\ln{[\left(1 - e^{-2 \tq}\frac {\left (1- \tq\tda\right)}
{\left (1+ \tq\tda \right)}\frac{\left (1 - \tq\tdo \right)}{\left (1 + \tq\tdo\right)}\right)]} + \nonumber \\
 & &\ln{[\left(1 - e^{-2 \tq}\frac {\left (\tq - \tda\xi^2\right)}
{\left (\tq + \tda\xi^2\right)}\frac{\left (\tq - \tdo \xi^2 \right)}{\left (\tq + \tdo \xi^2 \right)}\right)]}
 \label{eq:ene1}
\eeqa
with $p=a q,\tilde{k}_{(x,y)}=a k_{(x,y)},
\xi=a\zeta, \tilde{\delta}_{(a,0)}=\frac{\delta_{(a,0)}}{a}$.

In the following we will concentrate on the Casimir force, it can be written:
\beqa
F_R &=& -\frac{1}{2 \pi^2 a^4}\int_0^\infty d\tz\int_{\tz}^\infty \tq^2 d\tq
\frac{ e^{-2  \tq}  (\tdo \tq-1) (\tda \tq-1)}{(\tdo \tq+1)
   (\tda \tq+1) \left(1-\frac{e^{-2 \tq} (\tdo \tq-1) (\tda
   \tq-1)}{(\tdo \tq+1) (\tda \tq+1)}\right)}+ \nonumber \\
& &\frac{ e^{-2 \tq}
   \left(\tq-\tdo \tz^2\right) \left(\tq-\tda \tz^2\right)}{\left(\tdo
   \tz^2+\tq\right) \left(\tda \tz^2+\tq\right) \left(1-\frac{e^{-2 \tq} \left(\tq-\tdo \tz^2\right) \left(\tq-\tda \tz^2\right)}{\left(\tdo \tz^2+\tq\right) \left(\tda
   \tz^2+\tq\right)}\right)}
 \label{eq:for1}
\eeqa
Now it is not difficult to show that the contribution coming from
the point $\tz=0$ is zero, thus we can safely remove this point from
the integral, which allows us to rewrite the integral:
\beqa
F_R &=& -\frac{1}{2 \pi^2 a^4}\int_0^\infty d\tz\int_{\tz}^\infty \tq^2 d\tq
\sum_{n=1}^\infty e^{-2\tq n}  \left[\frac{(\tdo \tq-1) (\tda \tq-1)}{(\tdo \tq+1)
   (\tda \tq+1) }\right]^n+ \nonumber \\
& & e^{-2\tq n} \left[\frac{
   \left(\tq-\tdo \tz^2\right) \left(\tq-\tda \tz^2\right)}{\left(\tdo
   \tz^2+\tq\right) \left(\tda \tz^2+\tq\right)}\right]^n \nonumber \\
& & =:  \sum_{n=1}^\infty F^{~n}_R
 \label{eq:for2}
\eeqa
The absolute values of the terms in the brackets are always less or
equal to one, in the case $\tdo=\tda=(0,\infty)$ they are maxima
$(1)$ and we have:
\beqa
F_R &=& -\sum_{n=1}^\infty \frac{1}{ \pi^2 a^4}\int_0^\infty d\tz\int_{\tz}^\infty \tq^2 d\tq  e^{-2 \tq n}=-\sum_{n=1}^\infty \frac{1}{ \pi^2 a^4}\int_0^\infty d\tz
\frac{e^{-2n \tz} (2  n \tz ( n \tz+1)+1)}{4 n^3} \nonumber \\
&=& -\sum_{n=1}^\infty \frac{3}{8 a^4 n^4 \pi ^2}=-\frac{\pi ^2}{240 a^4}.
 \label{eq:for0}
\eeqa
In contrast, if we take $\tdo=\infty,~\tda=0$ or viceversa they take on minimum values of $(-1)$ and we obtain
\beqa
F_R &=& -\sum_{n=1}^\infty \frac{1}{ \pi^2 a^4}\int_0^\infty d\tz\int_{\tz}^\infty \tq^2 d\tq (-1)^n e^{-2 \tq n}= \nonumber \\
&=& -\sum_{n=1}^\infty \frac{3(-1)^n }{8 a^4 n^4 \pi ^2}=
\frac{7}{8}\frac{\pi ^2}{240 a^4}.
 \label{eq:forinf}
\eeqa
Thus, in this case we recover the result obtained by Boyer
\cite{boye74} for two non dispersive mirrors having
$\epsilon=(\infty,1), \mu=(1,\infty)$ respectively, see also
\cite{henjou,astir08}. The upper calculation also constitutes an independent demonstration of the result found by Henkel and
Joulain eq.(4) of \cite{henjou}.

From eq. (\ref{eq:for2}) we may understand intuitively
what kind of conditions must be satisfied to have repulsion. Indeed,
if the two slabs are made of the same material we have $\tda=\tdo$
and the expression of the force becomes:
\bee
F_R =-\sum_{n=1}^\infty \frac{1}{2 \pi^2 a^4}\int_0^\infty d\tz\int_{\tz}^\infty \tq^2 d\tq e^{-2\tq n} \left[\left( \frac{ \tdo \tq-1}{\tdo \tq+1}\right)^{2 n}+\left( \frac{ \tq-\tdo \tz^2}{\tdo \tz^2+\tq }\right)^{2 n}\right].
\ee
In this case the integrand is always positive and the force will be
always attractive. The only possibility to have repulsion is to have
$\tda\neq\tdo$, such that $ \left[\frac{(\tdo \tq-1) (\tda
\tq-1)}{(\tdo \tq+1) (\tda \tq+1)}\right]^n+
\left[\frac{\left(\tq-\tdo \tz^2\right) \left(\tq-\tda
\tz^2\right)}{\left(\tq+\tdo \tz^2\right) \left(\tq+\tda
\tz^2\right)}\right]^n$ be negative. Fortunately, the series starts
with the term $n=1$ so that the possibility is not ruled out.

In the next section we will study the case $\tda\ll1$ and we will
concentrate on the first term of the series: $n=1$.

\section{Sufficient Conditions for Repulsion}

In the following we will develop $F^1_R$ at first order around
$\tda=0$. After the integration on the $\tq$ variable we will study
the behavior of the remaining integrand which will be a function of
$\tz$ only and determine what conditions must be satisfied to have
repulsion.
\beqa
F^1_R &=& \frac{1}{2 \pi^2 a^4}\int_0^\infty d\tz\int_{\tz}^\infty \tq^2 d\tq e^{-2\tq } \left[(2\tq\tda-1)\left( 1-\frac{2\tdo \tq}{\tdo \tq+1}\right)+
\left( 2\tda\frac{\tz^2}{\tq}-1\right)\left(1- \frac{ 2\tdo \tz^2}{\tdo \tz^2+\tq }\right)\right]\nonumber \\
&=& \frac{1}{8 \pi^2 a^4}\int_0^\infty d\tz \frac{e^{-2\tz}}{\tdo^3}\left\{
I_1(\tz)-8(1+2\tda)I_2\left(2\tz+\frac{2}{\tdo}\right)+8\tdo^5(\tdo+2\tda)\tz^6I_2\left(2\tz+
2{\tdo \tz^2}\right)\right\} \nonumber \\
& =:& \frac{1}{128 \pi^2 a^4}\int_0^\infty d\tz f^1_R(\tz) \label{eq:i123}
\eeqa
With:
\beqa
I_1(\tz) &=& -4 \tz^4 \tdo^5+\left(4 \tz^3+2 \tz^2-2 \tda \left(2 \tz (\tz+1)
   \left(2 \tz^2+3\right)+3\right)\right) \tdo^4+ \nonumber \\
& &\tda (2 \tz (4 \tz(\tz+2)+7)+7) \tdo^3+(\tda (-8 \tz-4)-4 \tz-2) \tdo^2+(8
   \tda+4) \tdo\nonumber \\
I_2(\tz) &=&  e^{\tz} E_1(\tz)\nonumber
\eeqa
$E_n(x)$ is the {\em exponential integral function} \cite{abste}.
Let us study $f^1_R(\tz)$ for the two regimes,
$0\leq\tda<\tdo\ll1$ and $0\leq\tda\ll1,\tdo\gg1$. In the first case
we find:
\bee
F^1_R=\frac{1}{8\pi^2 a^4}\int d\tz e^{-2\tz}\left[{(\tdo+\tda)} \left( 8 \tz^3+8 \tz^2+6 \tz+3\right)- \left(2 \tz^2+2\tz+1\right)\right]
\ee
Thus, in the range of frequencies relevant to
the Casimir effect, $\tz\sim1$, the condition for having repulsion is, at second order in $\tz$:
\bee
{\tdo} >\frac{2 + 4 \tz (1 + \tz)}{3 + 2 \tz (3 + 4 \tz (1 + \tz))}-\tda
\sim-\tda + 0.775- 0.494 \tz + 0.119 \tz^2.
\ee
This shows that it would be possible to have repulsion if $\tda \approx 0$ and $\tdo>0.4$. However this last condition is in
contradiction with the assumption $\tdo\ll1$. Let us also note
that if we assume $\tda=\tdo=const$ we can evaluate
all terms of the series (\ref{eq:for2}) and recover the
result of Mostepanenko and Trunov \cite{moste84}, (see
also \cite{saha02} for equivalent results for a scalar field).

If $0\leq\tda\ll1,\tdo\gg1$ we find at first order (for the
asymptotic expansion of $E_n(\xi)$ see \cite{abste})
\beqa
F^1_R &=&\frac{1}{16 \pi^2 a^4}\int d\tz \frac{e^{-\tz}}{\tdo\tz^2}\left\{-\tdo^2 A(\tda,\tz)+\tdo B(\tda,\tz)+C(\tda,\tz)\right\} \label{eq:ff1r}
\eeqa
with:
\begin{eqnarray*}
A(\tda,\tz) &=&-2 \tda \tz^2 (3 + 2 \tz (3 + \tz (3 + 2 \tz))) \\
B(\tda,\tz) &=&2 \tz^2 + 4 \tz^3 (1 + \tz) + \tda \tz^2 (7 + 2 \tz (7 + 6 \tz)) \\
C(\tda,\tz) &=&-3 - 6 \tz - 8 \tz^2 - 8 \tz^3 + 8 \tda \tz^4
\end{eqnarray*}
The term within the curly brackets in Eq. (\ref{eq:ff1r}) is a second
order polynomial in $\tdo$ and, to have repulsion, it must be
positive. Since the coefficient of the $\tdo^2$ term is always
negative we must require that the discriminant of the associated
second order equation is positive. This discriminant is a second
order polynomial in $\tda$ and we have to study the associated
second order equation:
\bee
\tda^2 D(\tz)+\tda E(\tz)+F(\tz)=0 \label{eq:da}
\ee
with
\begin{eqnarray*}
D(\tz) &=& 49 \tz^4 + 196 \tz^5 + 556 \tz^6 + 720 \tz^7 + 528 \tz^8 +
  256 \tz^9 \\
E(\tz) &=& -72 \tz^2 - 288 \tz^3 - 596 \tz^4 - 848 \tz^5 -
  744 \tz^6 - 432 \tz^7 - 160 \tz^8 \\
F(\tz) &=&4 \tz^4 + 16 \tz^5 + 32 \tz^6 + 32 \tz^7 + 16\tz^8.
\end{eqnarray*}

Since $D(\tz)>0$, for the polynomial to be positive we have to
choose $\tda$ such that it lies outside the interval defined by the two roots $\tda^{min}$ and $\tda^{max}$ of the corresponding equation, that is $\tda<\tda^{min}$ or $\tda>\tda^{max}$ for  $G=E^2-4D F>0 $. If $G\leq0$
we may take any value for $\tda$. In that case the two roots coincide if $\tda=0$ and become imaginary of $\tda > 0$ excluding any physical solution. Fig.1 illustrates the behavior of the two roots as a function of imaginary frequency.

\begin{figure}[h]
\centering
  \includegraphics[width=10.0cm]{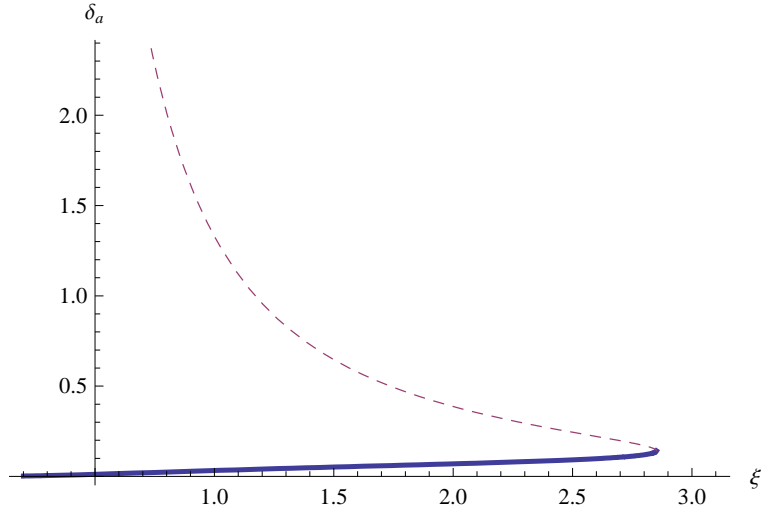}
  \caption{The two roots of eq. (\ref{eq:da}), $\tda^{min}$ and $\tda^{max}$, are shown as function of imaginary frequency $\xi$ (solid and dashed curve respectively). For $\tz\geq2.85$, $G\leq0$, at $\tz=2.85$ we have
$\tda^{min}=\tda^{max}$ and at larger frequency any value of $\tda$
will give rise to a positive value for (\ref{eq:da}). }
\end{figure}

Since $\tda^{max}$ can be larger than $1$ (see Fig.1) and we assumed
$\tda\ll1$ we remain with the only possibility $\tda<\tda^{min}$.
Around $\tz\sim1$ the condition for the positivity can be written:
\bee
\tda<\tda^{min}=\frac{-E-2 \sqrt{E^2-4D C}}{2D}\sim
-0.0135 + 0.05148\tz - 0.00533\tz^2. \label{eq:tdamin}
\ee
If this inequality is satisfied the force density will be repulsive
for those values of $\tdo$ which satisfy
\bee
\tdo^{min} \leq\tdo\leq\tdo^{max} \label{eq:d0mima},
\ee
$\tdo^{min}$ ( $\tdo^{max}$ ) being the smaller (larger)  roots of the  associated equation:
\beqa
-\tdo^2A(\tda,\tz)+\tdo B(\tda,\tz)+C(\tda,\tz) &=&0 \\
\tdo^{min} = \frac{- B +\sqrt{B^2-4 A C}}{2 A} &\sim &  a_1+\tda a_2 \nonumber \\
\tdo^{max} = \frac{-B-\sqrt{B^2-4 A C}}{2 A} &\sim&  a_3-\frac{1}{\tda}a_4-\tda a_2 \nonumber
\eeqa
where
\begin{eqnarray*}
a_1 &=& 10.48- 12.56 \tz + 4.58 \tz^2,~
a_2 = 152.438- 230.856 \tz + 93.118 \tz^2,\\
a_3 &=& -9.16122 + 12.0305\tz - 4.50083\tz^2,~
a_4 = 0.376294- 0.126549 \tz + 0.013413 \tz^2
\end{eqnarray*}
Note that in the case of an ideal mirror at $z=a$ we have $\tda=0$,
$\tdo^{max}\rightarrow\infty$ and the only condition to be satisfied
is:
$$\tdo>\tdo^{min}=10.48- 12.56 \tz + 4.58 \tz^2.$$
Thus the situation in which one mirror is ideal gives rise to quite
different results than the ones obtained when both are real. When both mirrors are real, they both must satisfy restrictions to ensure repulsion and, moreover,
a precise relation between the two penetration depths must be
fulfilled ( eqs. (\ref{eq:tdamin},\ref{eq:d0mima})).

\section{Discussion and Conclusions}

Let us apply our results to one known situation, that is of two
mirrors described by the plasma model. In this case we have
$$\tilde{\delta}_{(0,a)}=\frac{1}{\sqrt{(\xi^p_{(0,a)})^2+\xi^2}}~~\text{
with} ~~ \xi^p_{(0,a)}=a \omega^p_{(0,a)}$$ $\omega^p_{(0,a)}$ being
the plasma frequency of the mirror in $z=(0,a)$ respectively.
Condition (\ref{eq:tdamin}) gives $\xi^p_{a}>\frac{1}{0.032}\sim30.65$ which means that the mirror in $a$ must have a plasma frequency $\omega^p_{a}\geq30.65/a$, but condition (\ref{eq:d0mima}) implies $ \frac{1}{\sqrt{(\xi^p_{0})^2+1}}>4.37 $ which, being $\sqrt{(\xi^p_{0})^2+1}\geq1$, is impossible.

Let us consider now the case of hypothetical
materials having $\tdo(\xi)=k/\xi$ with $k=1,2,3,4$. There we obtain for the Casimir force, using the exact first three
terms of the series eq.(\ref{eq:for2}):
\begin{eqnarray*}
a^4 F_R(k=1) &=& -0.0007-0.0004-0.0000=-0.0011\\
a^4 F_R(k=2) &=& 0.0084 -0.0005+0.0000=0.0079\\
a^4 F_R(k=3) &=&0.0133-0.0006+0.0001=0.0128\\
a^4 F_R(k=4) &=&0.0164 -0.0008+ 0.0001=0.0157
\end{eqnarray*}
The result is illustrated on Fig. 2. The left hand part shows $\tda^{min}$ (dashed line) and $\tda=\frac{1}{\sqrt{30.65^2+\xi^2}}$ (solid line) as a function of imaginary frequency. In the right hand part the shaded area gives the range of values of $\tdo$ given by condition (\ref{eq:d0mima}) for which the Casimir force becomes repulsive while the dashed, dotted, dotted-dashed and solid lines, decreasing monotonously with increasing frequency, represent $\tdo$ for $k=1,2,3,4$ respectively. For $k=2,3,4$ the force turns out to be the more repulsive the higher the $k$-value, even though the values of $\tdo$ are only on the limit of the favorable region.

\begin{figure}[h]
\centering
  \includegraphics[width=7.0cm]{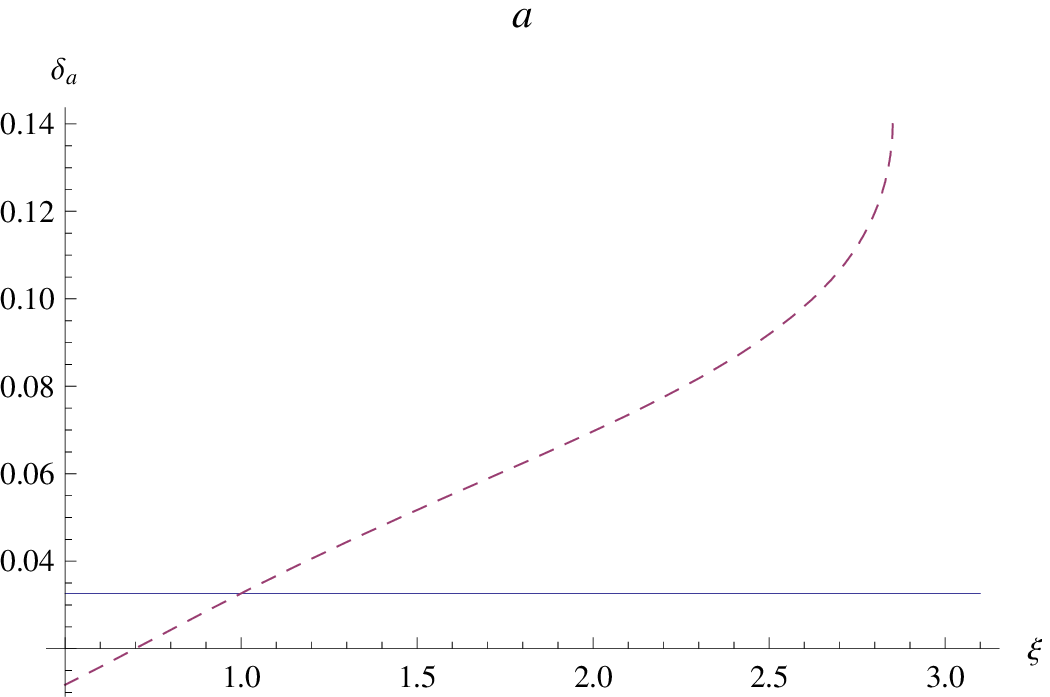}
    \includegraphics[width=7.0cm]{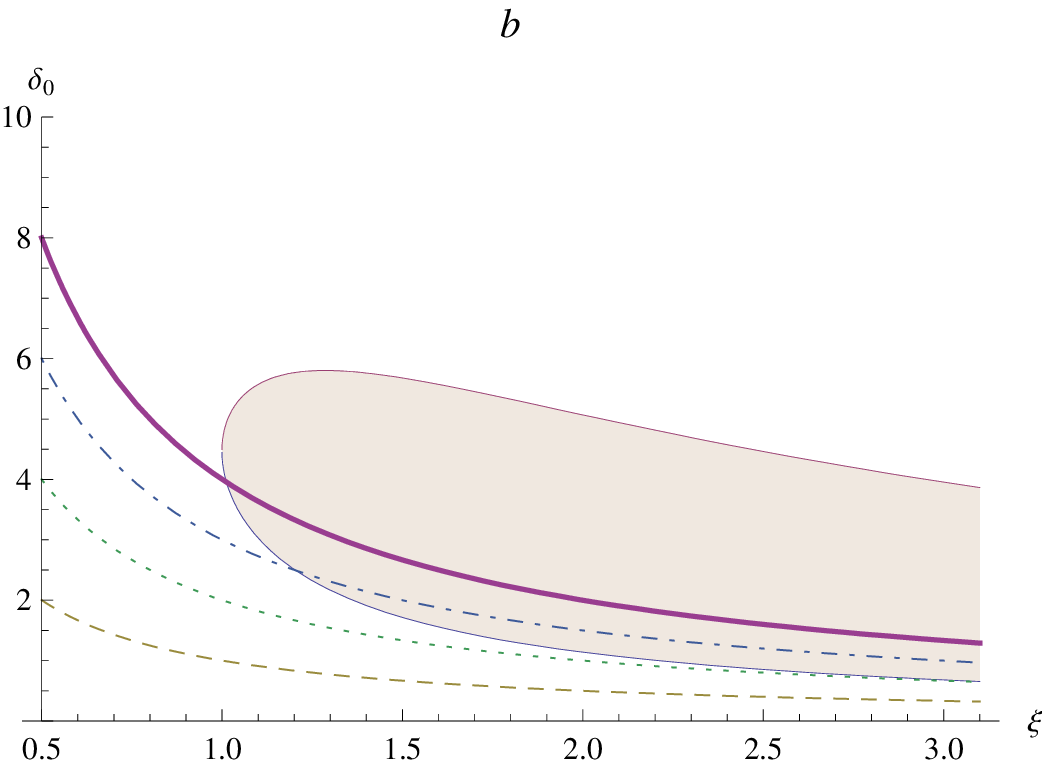}
\caption{In $a$  $\tda^{min}$ and $\tda=\frac{1}{\sqrt{30.65^2+\xi^2}}$ are shown as dashed and solid curves respectively.
In $b$ the area between the two curves
$\tdo^{max},\tdo^{min}$ (shaded area) and the curves $k/\xi$ for
$k=1,2,3,4$ dashed, dotted, dash-dotted, and thick respectively are shown.
For $\xi<1$ $\tda>\tda^{min}$ and consequently
the results are imaginary and no physical solution exists.}
\end{figure}

In conclusion we have found sufficient conditions for the Casimir force to be repulsive with an
approach considering only the skin depth and needing no specific model of dielectric or magnetic properties.
It would be interesting to study now how much
these conditions can be softened, as after all to have a positive integral it is
not necessary to have a positive integrand. From this point of view our analysis must be deepened trying
to obtain analytical necessary conditions for having a repulsive
force. Nonetheless our result demonstrate that repulsion is possible
if the penetration depth of the two mirrors satisfy appropriate
relations.

The approach seems promising
as it can be extended to anisotropic material
characterized by a tensorial surface impedance and to more general
material \cite{esqui03}.

It would also be very interesting to derive the skin depth at optical frequencies from the available
tabulated data to search for materials matching the conditions we
have established and to use our result to design new materials such as to have repulsive properties.

\subsection*{Acknowledgment}

Luigi Rosa thanks  University of Napoli Federico II for partial
financial support and \\ PRIN {\em Fisica Astroparticellare}. He
also thanks the Laboratoire Kastler Brossel for the kind
hospitality. The authors thank the ESF Research Networking Programme CASIMIR
(www.casimir-network.com) for providing excellent opportunities for
discussions on the Casimir effect and related topics.

\end{document}